\documentclass{article}
\usepackage[margin=3cm]{geometry}

\usepackage{amsmath,amssymb,amsfonts}
\usepackage{mathrsfs} 
\usepackage{mathtools}

\usepackage{graphicx,color}
\usepackage[nocompress]{cite}

\usepackage{authblk}

\newcommand{\ourG}{\mathbf{G}}
\newcommand{\ourB}{\mathbf{B}}

\usepackage{hyperref}
\hypersetup{colorlinks=true,allcolors=blue,urlcolor=cyan}

\pagecolor{white}

\begin{document}

\title{A new 2D formulation of modified General Relativity}

\author[1,2]{Christian G B\"ohmer\footnote{Email: c.boehmer@ucl.ac.uk}}
\author[1]{Erik Jensko\footnote{Email: erik.jensko.19@ucl.ac.uk}}
\affil[1]{Department of Mathematics, University College London, \authorcr Gower Street, London WC1E 6BT, UK\medskip}
\affil[2]{Astrophysics Research Centre, School of Mathematics, \authorcr Statistics and Computer Science, University of KwaZulu-Natal, \authorcr Private Bag X54001, Durban 4000, South Africa\medskip}

\date{\today} 

\maketitle

\begin{abstract}
  It is well known that the Einstein-Hilbert action in two dimensions is topological and yields an identically vanishing Einstein tensor. Consequently one is faced with difficulties when formulating a non-trivial gravity model. We present a new, intrinsically two-dimensional, approach to this problem based on the Einstein action. This yields a well defined variational approach which results in new field equations that break diffeomorphism invariance. Our proposed approach does not require the introduction of additional scalar fields, nor the use of conformal transformations. However, we can show how including conformal counter terms leads to equivalent results. In doing so, we can provide an explanation for why previous approaches worked. Solutions to the field equations are briefly discussed.
\end{abstract}

\section{Introduction}

The Einstein field equations derived from the Einstein-Hilbert action in two dimensions (2D) are problematic since the Einstein tensor vanishes identically. This is because the action is topological in two dimensions, leading to the identity $R_{\rho \sigma}=g_{\rho \sigma} R/2$ which relates the Ricci tensor and the Ricci scalar. Many different approaches have been considered in the past to construct 2D gravitational toy models, see~\cite{Grumiller:2006rc} and references therein. Of particular interest for the present work is the approach used in~\cite{Mann:1992ar} where the authors used a conformal transformation to formulate an action in the limit $D \rightarrow 2$. Ideas along these lines have had a recent surge in interest when these were applied to what is now called four-dimensional Einstein-Gauss-Bonnet Gravity~\cite{Glavan:2019inb}, together with various comments and critiques~\cite{Hennigar:2020lsl,Ai:2020peo,Gurses:2020rxb}. Nonetheless this model has attracted a substantial amount of interest, see~\cite{Fernandes:2022zrq} for a recent review. The limiting theory of this model is a Horndeski type theory~\cite{Lu:2020iav}. There are also close connections between the equations presented below and 2D Jackiw-Teitelboim gravity~\cite{Jackiw:1984je,Teitelboim:1983ux}. Similar relations have also been found in other $D \rightarrow 2$ limits of gravity~\cite{Lemos:1993hz,Mao:2022zrf,Mann:1992ar}, which we refer to for comparison. An alternative approach is to look beyond the Riemannian framework, such as in Poincar\'{e} gravity~\cite{Mielke:1993nc,Obukhov:1997uc}.

Here will present a novel way to study 2D gravity within the standard Riemannian geometry. Our approach is intrinsically 2D and hence does not require the use of dimensional limits. Moreover, we do not need to introduce a scalar field via conformal transformations or otherwise, nor do we have to subtract certain counter terms to make things work. However, in our model we are also able to show how conformal transformations and scalar fields are related to our approach and provide an {\it a posteriori} justification for their use and the specific counter terms which have been used in previous works. This result gives an illuminating view of the role of conformal counter terms in these theories.

Our approach explicitly breaks the diffeomorphism symmetry present in standard gravitational theories. As a result of breaking this symmetry, additional non-trivial constraint equations are generated, replacing the usual Bianchi identity associated with the diffeomorphism invariance of General Relativity. These new constraints are coordinate-dependent, and we show how they can be satisfied by choosing appropriate coordinates for given solutions of the theory. Crucially, this allows interesting non-trivial solutions in two dimensions.

\section{The model and its key properties}

Let us begin by recalling the well-known fact that the Einstein field equations can be derived from the so-called Einstein action 
\begin{align}
  S_{\rm Einstein} = \frac{1}{2\kappa} \int
  g^{\mu \nu} (\Gamma^{\sigma}_{\lambda \mu} \Gamma^{\lambda}_{\sigma \nu} -
  \Gamma^{\sigma}_{\mu \nu} \Gamma^{\lambda}_{\sigma \lambda}) \sqrt{-g}\,d^4 x \,.
  \label{Einsteinaction}
\end{align}
This follows from the fact that this action differs from the usual Einstein-Hilbert action by a boundary term which does not contribute to the field equations. Specifically the Ricci scalar $R$ can be written as
  \begin{align}
    R = g^{\mu \nu} (\Gamma^{\sigma}_{\lambda \mu} \Gamma^{\lambda}_{\sigma \nu} -
    \Gamma^{\sigma}_{\mu \nu} \Gamma^{\lambda}_{\sigma \lambda}) + \ourB
  \end{align}
where $\ourB$ is the said boundary term, see~\cite{Boehmer:2021aji,Boehmer:2023fyl} for models of modified gravity based on the Einstein action and the boundary term. The Einstein action is not a true coordinate scalar; it is pseudo-invariant, in the sense that it is diffeomorphism invariant up to a boundary term. This can be explicitly seen by performing a general coordinate transformation and noting that the inhomogeneous terms take the form of a total derivative~\cite{Jensko:2023lmn}. In two dimensions, the action also takes the form of a total derivative, and a scalar-tensor limit can be obtained from the theory~\cite{Khodabakhshi:2022knu}.

We will now introduce the new object
\begin{align}
  \ourG :=  g^{\mu \nu} (c_1 \Gamma^{\sigma}_{\lambda \mu} \Gamma^{\lambda}_{\sigma \nu} -
  c_2 \Gamma^{\sigma}_{\mu \nu} \Gamma^{\lambda}_{\sigma \lambda} ) \,,
  \label{Gwithc}
\end{align}
where $c_1$ and $c_2$ are two arbitrary constants. The object $\ourG$ is not a true coordinate scalar unless $ c_1 = c_2 = 0$, in which case  $\ourG=0$, which follows from the fact that the Christoffel symbol is not a tensor. Moreover, it is no longer pseudo-invariant except when  $c_1 = c_2$. In the general case when $c_1 \neq c_2$, a general coordinate transformation introduces inhomogeneous terms that cannot be written in the form of a total derivative. The object $\ourG$ is therefore non-covariant and theories constructed from it will not be invariant under diffeomorphisms. The explicit breaking of this symmetry will lead to additional constraints, which will be shown in the following section.

Neglecting an overall scaling, the quantity $\ourG$ only depends on one constant. To see this, we can factor out the constant $c_1$, thereby introducing the ratio $c_1/c_2$ as the one free constant. As the overall scaling of the action will not affect the model under consideration, we will henceforth set $c_1=1$ and $c_2=\alpha$, with $\alpha$ being the unique parameter of the model
  \begin{align}
    \ourG = g^{\mu \nu} (\Gamma^{\sigma}_{\lambda \mu} \Gamma^{\lambda}_{\sigma \nu} -
  \alpha \, \Gamma^{\sigma}_{\mu \nu} \Gamma^{\lambda}_{\sigma \lambda} ) \, .
  \end{align}

Next, we introduce the new modified action in two dimensions
\begin{align}
  S_{c} = \frac{1}{2\kappa} \int \ourG \sqrt{-g}\,d^2 x \,,
  \label{actionc}
\end{align}
where $\kappa=8\pi G_{(2)}$ is the gravitational coupling constant and we use $G_{(2)}$ to denote Newton's 2D constant.
When considering variations with respect to the metric, this action will produce non-trivial field equation whenever $\alpha \neq 1$. When $\alpha = 1$ one would find the 2D Einstein tensor which vanishes identically. Note that this model is intrinsically two-dimensional as no reference to a higher-dimensional theory is made, nor are any additional fields present. 

One can now consider the following action
\begin{align}
  S = \lim_{\alpha \rightarrow 1} \frac{1}{2\kappa} \int \frac{\ourG}{1-\alpha} \sqrt{-g}\,d^2x +
  S_{\rm matter} \,,
  \label{actionfinal}
\end{align}
and study its metric variations, which will be well-defined in the limit $\alpha \rightarrow 1$. This is a straightforward exercise and gives
\begin{multline}
  \lim_{\alpha \rightarrow 1} \frac{1}{1-\alpha}\Big[
    \big( R_{\rho \sigma} + \partial_{(\rho} \Gamma_{\sigma)\lambda}^{\lambda} - \frac{1}{2} g_{\rho \sigma} g^{\eta \nu} \Gamma_{\mu \nu}^{\gamma} \Gamma^{\mu}_{\gamma \eta} \big) \\ +
    \alpha \big( - \frac{1}{2}g_{\rho \sigma} R - \partial_{(\rho} \Gamma_{\sigma)\lambda}^{\lambda} + \frac{1}{2} g_{\rho \sigma} g^{\eta \nu} \Gamma_{\mu \nu}^{\gamma} \Gamma^{\mu}_{\gamma \eta}\big)
    \Big] = \kappa T_{\rho \sigma} \,,
  \label{field0}
\end{multline}
In two dimensions we have the geometrical identity $R_{\rho \sigma}=g_{\rho \sigma} R/2$, which implies the field equations reduce to
\begin{align}
  \lim_{\alpha \rightarrow 1} \frac{1}{1-\alpha} \Bigl[(1-\alpha)
  \bigl(\frac{1}{2}Rg_{\rho \sigma} + \partial_{(\rho} \Gamma_{\sigma)\lambda}^{\lambda} - \frac{1}{2} g_{\rho \sigma} g^{\eta \nu} \Gamma_{\mu \nu}^{\gamma} \Gamma^{\mu}_{\gamma \eta}\bigr)\Bigr] = \kappa T_{\rho \sigma} \,.
  \label{field1}
\end{align}
It is now meaningful to take the limit $\alpha \rightarrow 1$ and to state the final field equations of this model
\begin{align}
  E_{\rho\sigma} := \partial_{\rho} \Gamma_{\sigma\lambda}^{\lambda} + \frac{1}{2}g_{\rho \sigma}
  \Bigl(R-g^{\eta \nu} \Gamma_{\mu \nu}^{\gamma} \Gamma^{\mu}_{\gamma \eta}\Bigr) =
  \kappa T_{\rho \sigma} \,.
  \label{field2}
\end{align}
The symmetrisation brackets in the connection term have been dropped since $\Gamma_{\sigma\lambda}^\lambda = \partial_{\sigma} \log \sqrt{-g}$ so that $\partial_{\rho} \Gamma_{\sigma\lambda}^{\lambda} = \partial_{\rho} \partial_{\sigma} \log \sqrt{-g}$ is automatically symmetric. 

 The final result of this derivation can be understood easily: the first part gives rise to \textit{half} of the Einstein tensor plus additional non-covariant terms, whilst the $\alpha$ part gives rise to the other \textit{half} and the same additional non-covariant terms with an overall minus sign. The usual limit of $\alpha \rightarrow 1$ gives the Einstein tensor with the additional terms mutually cancelling. However, the peculiar division by $(1-\alpha)$ gives something new, not previously studied. This result is unique to two dimensions, as in any other dimension the limit would give a singular contribution coming from the Einstein tensor. Consequently, one would have to deal with these singular terms in one way or another. Many of the previously mentioned approaches introduced scalar fields to formulate dimensionally reduced theories. Due to our geometrical formulation intrinsic to 2D, there is no natural scalar present. Moreover, no counter terms have to be introduced since our field equations are regular in the limit.

 However, we can rewrite our field equation slightly by viewing the determinant of the metric as a scalar field, and this will later be linked back to conformal transformations. To do so we set $\Phi:=\log\sqrt{-g}$ and consider the trace of~(\ref{field2}). One finds
  \begin{align}
    g^{\mu\nu} \partial_{\mu} \partial_{\nu} \Phi +
    (R-g^{\eta \nu} \Gamma_{\mu \nu}^{\gamma} \Gamma^{\mu}_{\gamma \eta}) = \kappa T \,,
  \end{align}
  where $T=g^{\rho\sigma}T_{\rho\sigma}$ is the trace of the energy-momentum tensor. We also have the following form of the Ricci scalar
  \begin{align}
    R = e^{-\Phi}g^{\mu\nu}\partial_\alpha(e^{\Phi} \Gamma^\alpha_{\mu\nu}) -
    g^{\mu\nu} \partial_{\mu} \partial_{\nu} \Phi -
    g^{\mu\nu}\Gamma^\alpha_{\mu\beta} \Gamma^\beta_{\nu\alpha}\,,
  \end{align}
which one could use to further rewrite the field equations. We can now decompose the field equations into a trace-free equation and a trace equation respectively as
\begin{align}
  \partial_{\rho}\partial_{\sigma} \Phi - \frac{1}{2}g_{\rho\sigma}
  (g^{\mu\nu} \partial_{\mu} \partial_{\nu}\Phi) &=
  \kappa \Bigl(T_{\rho\sigma} -\frac{1}{2}g_{\rho\sigma} T\Bigr)\,,
  \label{field3a} \\
  g^{\mu\nu} \partial_{\mu} \partial_{\nu} \Phi +
  (R-g^{\eta \nu} \Gamma_{\mu \nu}^{\gamma} \Gamma^{\mu}_{\gamma \eta}) &=
  \kappa T \,.
  \label{field3b}
\end{align}
One could replace the two partial derivatives with covariant ones by also introducing compensating connection terms. However, this does not seem to improve the conceptual understanding of the equations. In contrast to~\cite{Mann:1992ar}, for example, these equations do not decouple in the following sense. One cannot isolate the scalar field, gravity or matter in such a way that its evolution becomes independent of the other.

Let us now consider the conformal transformation $\tilde{g}_{\mu\nu} = e^{2\phi} g_{\mu\nu}$, where a tilde will always denote the quantity in the conformal frame. A direct calculation gives the interesting result
\begin{align}
  \sqrt{-\tilde{g}}\tilde{\ourG} - \sqrt{-g}\ourG =
  (1-\alpha)\sqrt{-g}g^{\alpha\beta}\Gamma^\sigma_{\alpha\beta}\partial_\sigma \phi \,,
  \label{conf1}
\end{align}
which means that upon division by $(1-\alpha)$ one can derive, once again, a well-defined limiting theory when $\alpha \rightarrow 1$. Let us therefore define the following action with this additional regularisation counter term
\begin{align}
  S' = \lim_{\alpha \rightarrow 1} \frac{1}{(1-\alpha)}
  \int (\sqrt{-\tilde{g}}\tilde{\ourG} - \sqrt{-g}\ourG) d^2x =
  \int g^{\alpha\beta}\Gamma^\sigma_{\alpha\beta}\partial_\sigma \phi \sqrt{-g}\, d^2x \,.
  \label{conf2}
\end{align}
This takes the form of an unusual gravitational theory with non-minimally coupled scalar field. Containing an explicit connection term, this action is clearly not invariant under infinitesimal coordinate transformations. Let us thus treat action $S'$ as the starting of a theoretical model and consider independent variations of the scalar field $\phi$ and the metric tensor $g_{\alpha\beta}$. This leads to what is perhaps a most surprising result: the field equations when varying $\phi$ are identical to the trace of our field equation~(\ref{field2}) while variations with respect to the metric yield the trace-free equation~(\ref{field3a}). It needs to be emphasised here that the scalar field of the conformal transformation enters the theory as a true independent field while the previously introduced $\Phi$ is related to the determinant of the metric and thus carries no independent meaning.

One can somewhat reconcile these observations when going back to~(\ref{Gwithc}). The $\alpha$ term is given by
  \begin{align}
    g^{\mu\nu} \Gamma^{\sigma}_{\mu \nu} \Gamma^{\lambda}_{\sigma \lambda} =
    g^{\mu\nu} \Gamma^{\sigma}_{\mu \nu} \partial_\sigma \log \sqrt{-g} =
    g^{\mu\nu} \Gamma^{\sigma}_{\mu \nu} \partial_\sigma \Phi \,,
  \end{align}
and hence is form equivalent to the term in the action~(\ref{conf2}). Since the first term and the $\alpha$ term both yield the same field equations in 2D, see~(\ref{field0}), it now clear that $S'$ also must lead to those same field equations. From our point of view it is most exciting to see that we can demonstrate that the conformal approach used previously can be well understood in an intrinsically 2D setting without having to worry about limits. We view this as the {\it a posteriori} justification of why previous approaches have worked. This approach can likely be extended to four-dimensional Einstein-Gauss-Bonnet Gravity.

\section{Conservation equations and exact solutions}

Before studying the field equations explicitly, we will first consider infinitesimal coordinate transformations of action~(\ref{actionc}). This allows us to investigate the consequences of breaking diffeomorphism invariance in our modified action. The object $\ourG$, which can be seen as the Lagrangian of our action, has the following transformation properties under infinitesimal coordinate transformations
\begin{align}
  \hat{\ourG}(\hat{x}) =
  \ourG - M^{\alpha \beta}{}_{\gamma} \partial_{\alpha} \partial_{\beta} \xi^{\gamma} \,,
  \quad
  M^{\alpha \beta}{}_{\gamma} &= 2 g^{\mu (\alpha} \Gamma^{\beta)}_{\mu \gamma} -
  \alpha g^{\alpha \beta} \Gamma^{\lambda}_{\lambda \gamma} -
  \alpha g^{\mu \nu} \delta^{(\beta}_{\gamma} \Gamma^{\alpha)}_{\mu \nu} \,,
  \label{G_infinitesimal}
\end{align}
which is in complete analogy to the results in~\cite{Boehmer:2021aji}. Following the same approach one finds that the change in the action generated by the vector field $\xi$ is
\begin{align}
  \delta_{\xi} S_c = \int \partial_{\alpha} \partial_{\beta}
  (\sqrt{-g} M^{\alpha \beta}{}_{\gamma} ) \xi^{\gamma} d^2 x \,,
  \label{diff1}
\end{align}
up to boundary terms. When $\alpha=1$ this becomes the twice-contracted Bianchi identity in non-standard form. This reflects the fact that the twice-contracted Bianchi identity is associated with the diffeomorphism invariance of General Relativity. However, when $\alpha \neq 1$ one arrives at
\begin{align}
  \partial_{\alpha} \partial_{\beta} (\sqrt{-g} M^{\alpha \beta}{}_{\gamma} ) =
  (1-\alpha) N_\gamma \,,
  \label{diff2}
\end{align}
with $N_\gamma$ given explicitly below. This is most interesting as it implies that one can again divide by $(1-\alpha)$ and arrive at some limiting conservation equation which has no obvious analogue to previously studied equations. It is also not clear at this point whether or not this limiting equation is meaningful.

One way to address this latter question is to compute $\nabla_\alpha E^\alpha_\gamma$, using the left-hand side of~(\ref{field2}) and using the standard way to compute the covariant derivative of a rank 2 object. Strictly speaking one should not do this as the covariant derivative of a connection is not well-defined and this left-hand side is not a rank 2 tensor. However, it turns out that the resulting equation is indeed equivalent to the above $N_\gamma$ and one finds $\nabla_\alpha E^\alpha_\gamma = N_\gamma$. Therefore, the field equations and the action both give rise to the same `conservation' equation which shows the internal consistency of this model. The explicit form reads
\begin{align}
  N_\gamma = \partial_\alpha E^\alpha_\gamma + \Gamma^\alpha_{\alpha\sigma} E^\sigma_\gamma -
  \Gamma^\sigma_{\alpha\gamma} E^\alpha_\sigma \,,
  \label{diff3}
\end{align}
with $E^{\alpha}_{\beta}$ given by the field equation~(\ref{field2}). Also see~\cite{Jackiw:2006kn}, where a similar result is obtained with diffeomorphism invariance being dynamically reinstated.

In the absence of matter it would be somewhat natural to seek only solutions satisfying $N_\gamma = 0$ while in the presence of matter this condition is not required, strictly speaking. When matter is present one could envisage a situation where $\nabla_\alpha T^\alpha_\gamma = -N_\gamma$ so that the \textit{total action} would remain invariant under diffeomorphisms, but the gravitational and matter actions would not be invariant independently. However, one can equally well argue for a different approach, namely ignoring $N_\gamma$ altogether. Since the field equations are non-tensorial one could simply neglect infinitesimal coordinate transformations. This could also be seen as some restrictions on the allowed coordinates of this theory~\cite{Jackiw:2007br}. When studying some explicit solutions of the field equations next, we will in fact find that $N_\gamma = 0$ for physically interesting solutions which means that these solution appear as if they were diffeomorphism invariant. It will be useful to include a cosmological constant $\Lambda$ in the field equations, so in the following we will work with
\begin{align}
  \partial_{\rho} \Gamma_{\sigma\lambda}^{\lambda} + \frac{1}{2}g_{\rho \sigma}
  \Bigl(R-g^{\eta \nu} \Gamma_{\mu \nu}^{\gamma} \Gamma^{\mu}_{\gamma \eta}\Bigr) -
  \Lambda g_{\rho \sigma} =
  \kappa T_{\rho \sigma} \,.
  \label{field2a}
\end{align}
The sign of the $\Lambda$ term was chosen so that $\Lambda>0$ corresponds to de Sitter space, as will be seen. Let us now discuss two interesting types of solutions. Firstly, we consider the cosmological metric
\begin{align}
  ds^2 = - dt^2 + a^2(t) dx^2 \,,
  \label{cos1}
\end{align}
where the lapse function was set to one. This line element gives $N_\gamma = 0$ which means that this class of solutions satisfies the conservation equation. The field equations with a cosmological term are
\begin{align}
  \frac{\dot{a}^2}{2a^2} - \Lambda = 0 \,, \qquad
  \frac{\ddot{a}}{a} - \frac{\dot{a}^2}{2a^2} - \Lambda = 0 \,,
\end{align}
which have the solution $a(t) = a_0 \exp(\pm \sqrt{2\Lambda} t)$, where $a_0$ is the constant of integration. This is the usual de Sitter solution with Ricci scalar $R = 4 \Lambda$, see~\cite{Sikkema:1989ib}, justifying the previous choice of sign in Eq.~(\ref{field2a}).

Secondly, turning our interests to black hole solutions, let us therefore consider the line element
\begin{align}
  ds^2 = - e^{2A(x)} dt^2 + e^{2B(x)} dx^2 \,.
  \label{bh1}
\end{align}
The vacuum field equations with (negative) cosmological constant $\Lambda=-1/\ell^2$ are given by
\begin{align}
  A'' + \frac{1}{2}(A'^2+B'^2) - A'B' - \frac{1}{\ell^2} e^{2B} &= 0 \,,\\
  B'' - \frac{1}{2}(A'^2+B'^2) + A'B' + \frac{1}{\ell^2} e^{2B} &= 0 \,,
  \label{bh2}
\end{align}
where one immediately notes that summing both equations yields $A''=-B''$. This implies the result $A=-B + b_1 + b_2 x$. Since the line element is static, we can always rescale the time coordinate to eliminate the constant $b_1$. This means we are free to choose $b_1=0$. Substituting this result into either of the two field equations gives
  \begin{align}
    B'' -2B'^2 + 2b_2 B' - \frac{(b_2)^2}{2} + \frac{1}{\ell^2} e^{2B} = 0\,,
  \end{align}
and can be solved explicitly. We find two types of solutions
\begin{alignat}{2}
  e^{-2B} &= \frac{x^2}{\ell^2} +  c_1 x + c_2 \,, &
  \qquad \text{if} \qquad b_2 &= 0\,,
  \label{eqsol1}\\
  e^{-2B} &= \frac{2}{(b_2^2)\ell^2} + e^{-b_2 x}(c_3  x + c_4)\,, &
  \qquad \text{if} \qquad b_2 &\neq 0\,.
  \label{eqsol2}
\end{alignat}
Here the $c_i$ are constants of integration. It is clear that these two types of solutions are distinct.

For the $b_2=0$ solution, setting the other integration constant to zero, $b_1=0$, leads to the simple relation between the metric functions $A(x)=-B(x)$. Additionally setting the constant $c_1=0$ leads to the following convenient form
\begin{align}
  e^{-2B} = e^{2A} = \frac{x^2 - \mathcal{C}^2}{\ell^2}\,,
  \label{bh4}
\end{align}
where we have defined $c_2=-\mathcal{C}/\ell^2$ with $\mathcal{C}$ a new constant of integration.

The most remarkable feature of this solution is that this takes the form of the famous BTZ black hole solution~\cite{Banados:1992wn}. It satisfies $R=-2/\ell^2$ as expected for this solution. This choice of constants of integration also ensure that $e^{-2B} = e^{2A}$, as in the Schwarzschild solution, for example. We also note that $N_\gamma = 0$ for this solution and remark that these solutions are analogous to those discussed in~\cite{Mann:1991qp,Mann:1989gh}. This is perhaps unsurprising as, after all, two-dimensional geometries are rather simple as the Riemann curvature tensor has only one independent component.

Finally, let us briefly comment on the $b_2 \neq 0$ solution~(\ref{eqsol2}). It does not satisfy the often found condition $e^{-2B} = e^{2A}$, it also does not give rise to a constant Ricci scalar which is expected for a vacuum solution in the presence of a cosmological constant. Therefore we regard this solution as less interesting and conclude that $b_2=0$ is the natural choice.

\section{Conclusions}

Using the original Einstein action as a starting point, we were able to construct an intrinsically 2D approach to formulate a theory of gravity in two dimensions. This is achieved without introducing additional fields into the theory. Our model shares many features with previously studied approaches but also shows distinct features. In general, our model will break diffeomorphism invariance, however, in a somewhat subtle way. When considering black hole solutions or cosmological solutions, for example, these appear as if the were invariant, which can be related back to the particular choice of coordinates employed. It should be noted that diffeomorphism invariance could be formally restored via the Stueckelberg trick, explicitly introducing new degrees of freedom associated with the broken symmetries, again see~\cite{Jensko:2023lmn}. From this perspective, the above theory is simply gauge-fixed by construction, but still completely legitimate from a mathematical standpoint. It would then be interesting to use this approach in the future to determine the exact number of dynamical degrees of freedom of this model. In the formulation presented here, however, choosing appropriate coordinates ensures the consistency of the model and its solutions. Again, this can be viewed as a form of gauge fixing.

Interestingly, we found that the determinant of the metric takes on the role of an unusual scalar that is non-minimally coupled to gravity. This was further expanded by using conformal transformations, which showed the equivalence of both methods. While there are many possible approaches to construct a 2D gravity model, our approach appears to provide a genuinely new angle to an old problem. 

\subsection*{Acknowledgements}
Erik Jensko is supported by EPSRC Doctoral Training Programme (EP/R513143/1).

\bibliographystyle{jhepmodstyle}
\bibliography{references}

\end{document}